\documentclass{llncs}

\usepackage{amssymb,amsmath,amscd,latexsym}
\usepackage{algorithm}
\usepackage{luca}
\def\qed{\rule{0.4em}{1.4ex}}

\newcommand{\pat}{\omega}
\newcommand{\Paths}{\Omega}
\newcommand{\PA}{1}
\newcommand{\PB}{2}
\newcommand{\straa}{\sigma}
\newcommand{\Straa}{\Sigma}
\newcommand{\strab}{\pi}
\newcommand{\Strab}{\Pi}
\newcommand{\SA}{S_1}

\newcommand{\SB}{S_2}

\newcommand{\SR}{S_{P}}
\newcommand{\ovSR}{\ov{S}_P}

\newcommand{\gamegraph}{G}

\newcommand{\winsure}[1]{\langle \! \langle #1 \rangle \! \rangle_{\mathit{sure}}  }
\newcommand{\winas}[1]{\langle \! \langle #1 \rangle\! \rangle_{\mathit{almost}} }

\newcommand{\winval}[1]{\langle \! \langle #1 \rangle\! \rangle_{\mathit{val}} }
\newcommand{\va}{\winval{1}}
\newcommand{\vb}{\winval{2}}
\newcommand{\was}{\winsure{1}}
\newcommand{\waa}{\winas{1}}

\newcommand{\wbs}{\winsure{2}}
\newcommand{\wba}{\winas{2}}

\newcommand{\outcome}{\mathrm{Outcome}}
\newcommand{\Prb}{\mathrm{Pr}}
\newcommand{\Inf}{\mathrm{Inf}}
\newcommand{\Reach}{{\text{\textrm{Reach}}}}
\newcommand{\Parity}{{\mathrm{Parity}}}

\newcommand{\coParityCond}{{\text{\textrm{coParity}}}(p)}

\newcommand{\Pat}{\Omega}
\newcommand{\Buchi}[1]{\textrm{B\"uchi}(#1)}
\newcommand{\coBuchi}[1]{\textrm{coB\"uchi}(#1)}

\newcommand{\attr}{\mathit{Attr}}

\newcommand{\Exp}{\mathbb{E}}
\newcommand{\Nats}{\mathbb{N}}
\newcommand{\nats}{\mathbb{N}}

\newcommand{\set}[1]{\{\: #1 \:\}}
\newcommand{\seq}[1]{\langle #1 \rangle}
\newcommand{\obciach}{\upharpoonright}
\newcommand{\trans}{\delta}
\newcommand{\distr}{{\cal D}}

\newcommand{\PM}{\mathit{PM}}

\newcommand{\Aa}{{\cal A}}
\newcommand{\ov}{\overline}

\newcommand{\wh}{\widehat}

\newcommand{\slopefrac}[2]{\leavevmode\kern.1em
  \raise .5ex\hbox{\the\scriptfont0 #1}\kern-.1em
  /\kern-.15em\lower .25ex\hbox{\the\scriptfont0 #2}}
\newcommand{\half}{\slopefrac{1}{2}}

\begin{document}
\title{Probabilistic Systems with \\ LimSup and LimInf Objectives }
\author{
 Krishnendu Chatterjee\inst{1}%
 \and Thomas A.\ Henzinger\inst{1,2}%
}
\institute{
EECS, UC Berkeley, USA
\and
 EPFL, Switzerland \\
{\tt \{c\_krish,tah\}@eecs.berkeley.edu}
}

\maketitle

\newif
  \iflong
  \longfalse
\newif
  \ifshort
  \shorttrue

\thispagestyle{empty}

\begin{abstract}
We give polynomial-time algorithms for computing the values of Markov 
decision processes (MDPs) with limsup and liminf objectives. 
A real-valued reward is assigned to each state, and the value of an 
infinite path in the MDP is the limsup (resp.\ liminf) of all rewards along 
the path.
The value of an MDP is the maximal expected value of an infinite path that 
can be achieved by resolving the decisions of the MDP.  
Using our result on MDPs, we show that turn-based stochastic games with 
limsup and liminf objectives can be solved in NP $\cap$ coNP.
\end{abstract}

\section{Introduction}

A \emph{turn-based stochastic game} is played on a finite graph with three 
types of states:
in player-1 states, the first player chooses a successor state from a given 
set of outgoing edges;
in player-2 states, the second player chooses a successor state from a given 
set of outgoing edges;
and probabilistic states, the successor state is chosen according to a given 
probability distribution.
The game results in an infinite path through the graph.
Every such path is assigned a real value, and the objective of player~1 is to
resolve her choices so as to maximize the expected value of the resulting path,
while the objective of player~2 is to minimize the expected value.
If the function that assigns values to infinite paths is a Borel function (in 
the Cantor topology on infinite paths), then the game is determined \cite{Mar98}:
the maximal expected value achievable by player~1 is equal to the minimal 
expected value achievable by player~2, and it is called the \emph{value} of 
the game. 

There are several canonical functions for assigning values to infinite paths. 
If each state is given a reward, then the \emph{max} (resp.\ \emph{min}) 
functions choose the maximum (resp.\ minimum) of the infinitely many rewards 
along a path;
the \emph{limsup} (resp.\ \emph{liminf}) functions choose the limsup 
(resp.\ liminf) of the infinitely many rewards;
and the \emph{limavg} function chooses the long-run average of the rewards.
For the Borel level-1 functions \emph{max} and \emph{min}, as well as for the
Borel level-3 function \emph{limavg}, computing the value of a game is known 
to be in NP $\cap$ coNP \cite{LigLip69}.
However, for the Borel level-2 functions \emph{limsup} and \emph{liminf}, 
only special cases have been considered so far.
If there are no probabilistic states (in this case, the game is called 
\emph{deterministic}), then the game value can be computed in polynomial 
time using value-iteration algorithms \cite{ChaHen07b};
likewise, if all states are given reward 0 or~1 (in this case, 
\emph{limsup} is a B\"uchi objective, and \emph{liminf} is a coB\"uchi 
objective), then the game value can be decided in NP $\cap$ coNP \cite{CJH04}.
In this paper, we show that the values of general turn-based stochastic 
games with limsup and liminf objectives can be computed in NP $\cap$ coNP.

It is known that pure memoryless strategies suffice for achieving the value 
of turn-based stochastic games with limsup and liminf objectives \cite{GimZie}.
A strategy is \emph{pure} if the player always chooses a unique successor 
state (rather than a probability distribution of successor states);
a pure strategy is \emph{memoryless} if at every state, the player always 
chooses the same successor state.
Hence a pure memoryless strategy for player~1 is a function from player-1 
states to outgoing edges (and similarly for player~2).
Since pure memoryless strategies offer polynomial witnesses, our result 
will follow from polynomial-time algorithms for computing the values of 
Markov decision processes (MDPs) with limsup and liminf objectives.
We provide such algorithms.

An MDP is the special case of a turn-based stochastic game which contains 
no player-1 (or player-2) states.
Using algorithms for solving MDPs with B\"uchi and coB\"uchi objectives, we 
give polynomial-time reductions from MDPs with limsup and liminf objectives 
to MDPs with max objectives.
The solution of MDPs with max objectives is computable 
by linear programming, and the linear program for MDPs with max objectives 
is obtained by generalizing the linear program for MDPs with reachability
objectives.
This will conclude our argument.

\medskip\noindent{\bf Related work.}
Games with limsup and liminf objectives have been widely studied in game 
theory;
for example, Maitra and Sudderth~\cite{MaitraSudderth} present several 
results about games with limsup and liminf objectives.
In particular, they show the existence of values in limsup and liminf games 
that are more general than turn-based stochastic games, such as concurrent 
games, where the two players repeatedly choose their moves simultaneously 
and independently, and games with infinite state spaces.
Gimbert and Zielonka have studied the strategy complexity of games with 
limsup and liminf objectives:
the sufficiency of pure memoryless strategies for deterministic games was 
shown in~\cite{GZ05}, and for turn-based stochastic games, in~\cite{GimZie}.
Polynomial-time algorithms for MDPs with B\"uchi and coB\"uchi objectives
were presented in~\cite{luca-Thesis}, and the solution turn-based stochastic 
games with B\"uchi and coB\"uchi objectives was shown to be in 
NP $\cap$ coNP in~\cite{CJH04}.
For deterministic games with limsup and liminf objectives polynomial-time 
algorithms have been known, for example, the value-iteration algorithm 
terminates in polynomial time~\cite{ChaHen07b}.

\section{Definitions}
\label{section:definition}
We consider the class of turn-based probabilistic games and some of its subclasses.

\medskip\noindent{\bf Game graphs.}
A \emph{turn-based probabilistic game graph} 
(\emph{$2\half$-player game graph})
$\gamegraph =((S, E), (\SA,\SB,\SR),\trans)$ 
consists of a directed graph $(S,E)$, a partition $(\SA$, $\SB$,
$\SR)$ of the finite set $S$ of states, and a probabilistic transition 
function $\trans$: $\SR \rightarrow \distr(S)$, where $\distr(S)$ denotes the 
set of probability distributions over the state space~$S$. 
The states in $\SA$ are the {\em player-$\PA$\/} states, where player~$\PA$
decides the successor state; the states in $\SB$ are the {\em 
player-$\PB$\/} states, where player~$\PB$ decides the successor state; 
and the states in $\SR$ are the {\em probabilistic\/} states, where
the successor state is chosen according to the probabilistic transition
function~$\trans$. 
We assume that for $s \in \SR$ and $t \in S$, we have $(s,t) \in E$ 
iff $\trans(s)(t) > 0$, and we often write $\trans(s,t)$ for $\trans(s)(t)$. 
For technical convenience we assume that every state in the graph 
$(S,E)$ has at least one outgoing edge.
For a state $s\in S$, we write $E(s)$ to denote the set 
$\set{t \in S \mid (s,t) \in E}$ of possible successors.
The {\em turn-based deterministic game graphs} (\emph{2-player game graphs})
are the special case of the $2\half$-player game graphs with $\SR = \emptyset$.
The \emph{Markov decision processes} (\emph{$1\half$-player game graphs}) 
are the special case of the $2\half$-player game graphs with 
$\SA = \emptyset$ or $\SB = \emptyset$. 
We refer to the MDPs with $\SB=\emptyset$ as \emph{player-$\PA$} MDPs,
and to the MDPs with $\SA=\emptyset$ as \emph{player-$\PB$} MDPs.

\medskip\noindent{\bf Plays and strategies.}
An infinite path, or a \emph{play}, of the game graph $\gamegraph$ is an 
infinite 
sequence $\pat=\seq{s_0, s_1, s_2, \ldots}$ of states such that 
$(s_k,s_{k+1}) \in E$ for all $k \in \Nats$. 
We write $\Paths$ for the set of all plays, and for a state $s \in S$, 
we write $\Paths_s\subseteq\Paths$ 
for the set of plays that start from the state~$s$.
A \emph{strategy} for  player~$\PA$ is a function 
$\straa$: $S^*\cdot \SA \to \distr(S)$ that assigns a probability 
distribution to all finite sequences $\vec{w} \in S^*\cdot \SA$ of states 
ending in a player-1 state 
(the sequence represents a prefix of a play).
Player~$\PA$ follows the strategy~$\straa$ if in each player-1 
move, given that the current history of the game is
$\vec{w} \in S^* \cdot \SA$, she chooses the 
next state according to the probability distribution $\straa(\vec{w})$.
A strategy must prescribe only available moves, i.e., 
for all $\vec{w} \in S^*$,
$s \in \SA$, and $t \in S$, if $\straa(\vec{w} \cdot s)(t) > 0$, then 
$(s, t) \in E$.
The strategies for player~2 are defined analogously.
We denote by $\Straa$ and $\Strab$ the set of all strategies for player~$\PA$
and player~$\PB$, respectively.

Once a starting state  $s \in S$ and strategies $\straa \in \Straa$
and $\strab \in \Strab$ for the two players are fixed, the outcome
of the game is a random walk $\pat_s^{\straa, \strab}$ for which the
probabilities of events are uniquely defined, where an \emph{event}  
$\Aa \subseteq \Paths$ is a measurable set of plays. 
For a state $s \in S$ and an event $\Aa\subseteq\Paths$, we write
$\Prb_s^{\straa, \strab}(\Aa)$ for the probability that a play belongs 
to $\Aa$ if the game starts from the state $s$ and the players follow
the strategies $\straa$ and~$\strab$, respectively.
For a measurable function $f:\Paths \to \reals$ we denote by 
$\Exp_s^{\straa,\strab}[f]$ the \emph{expectation} of the function
$f$ under the probability measure $\Prb_s^{\straa,\strab}(\cdot)$.

Strategies that do not use randomization are called pure.
A player-1 strategy~$\straa$ is \emph{pure} if for all $\vec{w} \in S^*$
and $s \in \SA$, there is a state~$t \in S$ such that  
$\straa(\vec{w}\cdot s)(t) = 1$. 
A \emph{memoryless} player-1 strategy does not depend on the history of 
the play but only on the current state; i.e., for all $\vec{w},\vec{w'} \in 
S^*$ and for all $s \in S_1$ we have 
$\straa(\vec{w} \cdot s) =\straa(\vec{w}'\cdot s)$.
A memoryless strategy can be represented as a function 
$\straa$: $\SA \to \distr(S)$.
A \emph{pure memoryless strategy} is a strategy that is both pure and 
memoryless.
A pure memoryless strategy for player~1 can be represented as
a function $\straa$: $\SA \to S$.
We denote by 
$\Straa^{\PM}$ the set of pure memoryless strategies for player~1.
The pure memoryless player-2 strategies $\Pi^{\PM}$ are
defined analogously.

Given a pure memoryless strategy $\straa \in \Straa^{\PM}$, let 
$G_\straa$ be the game graph obtained from $G$ under the 
constraint that player~1 follows the strategy~$\sigma$.
The corresponding definition $G_\pi$ for a player-2 strategy 
$\pi\in\Pi^{\PM}$ is analogous, and we write $G_{\sigma,\pi}$ 
for the game graph obtained from $G$ if both players follow the 
pure memoryless strategies $\sigma$ and~$\pi$, respectively.
Observe that given a $2\half$-player game graph $G$ and a pure memoryless 
player-1 strategy~$\straa$, the result $G_{\straa}$ is a player-2 MDP. 
Similarly, for a player-1 MDP $G$ and a pure memoryless
player-1 strategy~$\straa$, the result $G_\straa$ is a Markov chain. 
Hence, if $G$ is a $2\half$-player game graph and the two players follow 
pure memoryless strategies $\straa$ and~$\strab$, the result
$G_{\straa,\strab}$ is a Markov chain.

\medskip\noindent{\bf Quantitative objectives.} A \emph{quantitative} objective
is specified as a measurable function $f:\Paths \to \reals$.
We consider \emph{zero-sum} games, i.e., games that are 
strictly competitive.
In zero-sum games the objectives of the players are functions $f$ and 
$-f$, respectively. 
We consider quantitative objectives specified as $\limsup$ and $\liminf$
objectives.
These objectives are complete for the second levels of the Borel hierarchy:
$\limsup$ objectives are $\Pi_2$ complete, and $\liminf$ objectives are 
$\Sigma_2$ complete.
The definitions of $\limsup$ and $\liminf$ objectives are as follows.
\begin{itemize}

\item 
\emph{Limsup objectives.} 
Let $r:S\to \reals$ be a real-valued reward function that assigns to 
every state $s$ the reward $r(s)$.
The \emph{limsup} objective $\limsup$ assigns to every play the maximum reward
that appears infinitely often in the play.
Formally, for a play $\pat=\seq{s_1,s_2,s_3,\ldots}$ we have
\[
\limsup(r)(\pat)=\limsup \langle r(s_i) \rangle_{i \geq 0}.
\]

\item \emph{Liminf objectives.} 
Let $r:S\to \reals$ be a real-valued reward function that assigns to 
every state $s$ the reward $r(s)$.
The \emph{liminf} objective $\liminf$ assigns to every play the maximum 
reward $v$
such that the rewards that appear eventually always in the play is at least $v$.
Formally, for a play $\pat=\seq{s_1,s_2,s_3,\ldots}$ we have
\[
\liminf(r)(\pat)=\liminf \langle r(s_i) \rangle_{i \geq 0}.
\]
The objectives $\limsup$ and $\liminf$ are complementary in the sense that for 
all plays $\omega$ we have 
$\limsup(r)(\pat)= -\liminf(-r)(\pat)$.
\end{itemize}
We also define the $\max$ objectives, as it will be useful in 
study of MDPs with $\limsup$ and $\liminf$ objectives.
Later we will reduce MDPs with $\limsup$ and $\liminf$ objectives to
MDPs with $\max$ objectives.
For a reward function $r:S \to \reals$ the \emph{max} objective 
$\max$ assigns to every play the maximum reward 
that appears in the play.
Observe that since $S$ is finite, the number of different rewards appearing
in a play is finite and hence the maximum is defined.
Formally, for a play $\pat=\seq{s_1,s_2,s_3,\ldots}$ we have
\[
\max(r)(\pat)=\max \langle r(s_i) \rangle_{i \geq 0}.
\]

\medskip\noindent{\bf B\"uchi and coB\"uchi objectives.} We define the
qualitative variant of $\limsup$ and $\liminf$ objectives, namely,
B\"uchi and coB\"uchi objectives.
The notion of qualitative variants of the objectives will be useful
in the algorithmic analysis of $2\half$-player games with $\limsup$ and
$\liminf$ objectives.
For a play $\pat$, we define $\Inf(\pat) = 
\set{s \in S \mid \mbox{$s_k = s$ for infinitely many $k \geq 0$}}$
to be the set of states that occur infinitely often in~$\pat$.
\begin{itemize}
\item
  \emph{B\"uchi objectives.}
  Given a set $B \subseteq S$ of B\"uchi states, the B\"uchi objective 
  $\Buchi{B}$ requires that some state in $B$ be visited
  infinitely often.
  The set of winning plays is 
  $\Buchi{B}= \set{\pat \in \Pat \mid 
  \Inf(\pat)\cap B\neq\emptyset}$.
 
\item \emph{co-B\"uchi objectives.}  
 Given a set $C \subseteq S$ of coB\"uchi states,
 the co-B\"uchi objective $\coBuchi{C}$ requires that only 
 states in $C$ be visited infinitely often.
  Thus, the set of winning plays is
$\coBuchi{C}=\set{\pat \in \Pat \mid 
  \Inf(\pat)\subseteq C}$.
  The B\"uchi and coB\"uchi objectives are dual in the sense that 
  $\Buchi{B}= \Pat \setminus \coBuchi{S\setminus B}$.
\end{itemize}
Given a set $B \subseteq S$, consider a boolean reward function $r_B$ such that
for all $s \in S$ we have $r_B(s)=1$ if $s \in B$, and $0$ otherwise.
Then for all plays $\pat$ we have $\pat \in \Buchi{B}$ iff 
$\limsup(r_B)(\pat)=1$.
Similarly, given a set $C \subseteq S$, consider a boolean reward function 
$r_C$ such that for all $s \in S$ we have $r_C(s)=1$ if $s \in C$, and $0$ otherwise.
Then for all plays $\pat$ we have $\pat \in \coBuchi{C}$ iff 
$\liminf(r_C)(\pat)=1$.

\medskip\noindent{\bf Values and optimal strategies.}
Given a game graph $G$, qualitative objectives $\Phi \subs \Paths$ for player~1 and 
$\Paths\setminus \Phi$ for player~2, and 
measurable functions $f$ and $-f$ for player~1 and player~2, 
respectively, we define the \emph{value} functions
$\va$ and $\vb$ for the players~1 and~2, respectively, as the following 
functions from the state space $S$ to the set $\reals$ of reals:
for all states $s\in S$, let
\[
\va^G(\Phi)(s)  =  
\displaystyle \sup_{\straa \in \Straa} \inf_{\strab \in \Strab} 
\Prb_s^{\straa,\strab}(\Phi); 
\]
\[
\va^G(f)(s)  = 
\displaystyle \sup_{\straa \in \Straa} \inf_{\strab \in \Strab} 
\Exp_s^{\straa,\strab}[f]; 
\]
\[
\vb^G(\Paths \setminus \Phi)(s)  =  
\displaystyle  \sup_{\strab\in \Strab} 
\inf_{\straa \in \Straa} \Prb_s^{\straa,\strab}(\Paths \setminus \Phi);
\]
\[ 
\vb^G(-f)(s)  =   
\displaystyle \sup_{\strab\in \Strab} 
\inf_{\straa \in \Straa} \Exp_s^{\straa,\strab}[-f].
\] 
In other words, the values $\va^G(\Phi)(s)$ and $\va^G(f)(s)$ give the maximal probability 
and expectation with which player~1 can achieve her objectives $\Phi$ and $f$ from state~$s$,
and analogously for player~2.
The strategies that achieve the values are called optimal:
a strategy $\straa$ for player~1  is \emph{optimal} from the state
$s$ for the objective $\Phi$ if 
$ \va^G(\Phi)(s) = \inf_{\strab \in \Strab} \Prb_s^{\straa, \strab}(\Phi)$;
and $\straa$ is \emph{optimal} from the state $s$ for $f$ if 
$\va^G(f)(s)=\inf_{\strab \in \Strab} \Exp_s^{\straa,\strab}[f]$.
The optimal strategies for player~2 are defined analogously.
We now state the classical determinacy results for $2\half$-player 
games with limsup and liminf objectives.

\begin{theorem}[Quantitative determinacy] 
\label{thrm:quan-det}
 For all $2\half$-player game graphs $G=((S,E),(S_1,S_2,\SR),\trans)$, the
following assertions hold.
\begin{enumerate}
\item
 For all reward functions $r:S \to \reals$ and all states~$s\in S$, we have 
 \[
  \va^G(\limsup(r))(s) + \vb^G(\liminf(-r))(s) =0; 
\]
\[
  \va^G(\liminf(r))(s) + \vb^G(\limsup(-r))(s) =0.
 \] 
\item Pure memoryless optimal strategies exist for both players from all 
states.
\end{enumerate}
\end{theorem}

The above results can be derived from the results in~\cite{MaitraSudderth};
a more direct proof can be obtained as follows: the existence of 
pure memoryless optimal strategies for MDPs with limsup and liminf 
objectives can be proved by extending the results known for B\"uchi
and coB\"uchi objectives.
The results (Theorem 3.19) of~\cite{Gimbert06} proved that if for a 
quantitative objective $f$ and its complement $-f$ pure memoryless optimal 
strategies exist in MDPs, then pure memoryless optimal strategies 
also exist in $2\half$-player games.
Hence the pure memoryless determinacy follows for $2\half$-player games
with limsup and liminf objectives.

\section{The Complexity of $2\half$-Player Games with \\ Limsup and LimInf Objectives}
In this section we study the complexity of MDPs and $2\half$-player games with 
limsup and liminf objectives.
We present polynomial time algorithms for MDPs and show that 
$2\half$-player games can be decided in 
NP $\cap$ coNP.
In the next subsections we present polynomial time algorithms 
for MDPs with limsup and liminf objectives by reductions to 
a simple linear-programming formulation, and then show
that $2\half$-player games can be decided in NP $\cap$ coNP.
We first present a remark and then present some basic results on MDPs.

\begin{remark}
Given a $2\half$-player game graph $G$ with a reward function $r:S \to 
\reals$ and a real constant $c$, consider the reward function $(r+c):
S \to \reals$ defined as follows:
for $s \in S$ we have $(r+c)(s)=r(s) +c$.
Then the following assertions hold: for all $s \in S$ 
\[
\va^G(\limsup(r+c))(s) =
\va^G(\limsup(r))(s)  +c;
\]
\[
\va^G(\liminf(r+c))(s) =
\va^G(\liminf(r))(s)  +c.
\]
Hence we can shift a reward function $r$ by a real constant $c$, and
from the value function for the reward function $(r+c)$, we can 
easily compute the value function for $r$.
Hence without loss of generality for computational purpose we assume 
that we have reward function with positive rewards, i.e.,
$r:S \to \reals^+$, where $\reals^+$ is the set of positive reals.
\end{remark}

\subsection{Basic results on MDPs}
In this section we recall several basic properties on MDPs. 
We start with the definition of \emph{end components} in 
MDPs~\cite{luca-Thesis,CY90} that play a role equivalent to closed recurrent 
sets in Markov chains.

\medskip\noindent{\bf End components.} Given an MDP 
$G=((S,E),(S_1,\SR),\trans)$, a set $U \subseteq S$ of states is an \emph{end 
component} if $U$ is $\trans$-closed (i.e., for all $s \in U \cap \SR$ we have
$E(s) \subseteq U$) and the sub-game graph of $G$ restricted to $U$ (denoted 
$G \obciach U$) is strongly connected.
We denote by $\cale(G)$ the set of end components of an MDP $G$. 
The following lemma states that, given any strategy (memoryless or not), with 
probability 1 the set of states visited infinitely often along a play is an
end component.
This lemma allows us to derive conclusions on the (infinite) set of plays in 
an MDP by analyzing the (finite) set of end components in the MDP.

\begin{lemma}\label{lemm:end-component}
\cite{luca-Thesis,CY90}
Given an MDP $G$, for all states $s \in S$ and all strategies 
$\straa \in \Straa$, 
we have 
$\Prb_s^{\straa}(\set{\pat \mid \Inf(\pat) \in \cale(G)})=1$.
\end{lemma}

For an end component $U \in \cale(G)$, consider the memoryless 
strategy $\straa_U$ that at a  state $s$ in $U \cap S_1$ plays 
all edges in $E(s) \cap U$ uniformly at random.
Given the strategy $\straa_U$, the end component $U$ is a closed connected 
recurrent set in the Markov chain obtained by fixing $\straa_U$.

\begin{lemma}\label{lemm:end-component1}
Given an MDP $G$ and an end component $U \in \cale(G)$, the strategy $\straa_U$
ensures that for all states $s \in U$, we have
$\Prb_s^{\straa_U}(\set{\pat \mid \Inf(\pat)=U})=1$.
\end{lemma}

\medskip\noindent{\bf Almost-sure winning states.}
Given an MDP $G$ with a B\"uchi or a coB\"uchi 
objective $\Phi$ for player~1, we denote by 
\[
W_1^G(\Phi) = \set{s \in S \mid \va(\Phi)(s)=1};
\]
the sets of states such that the values for player~1 is~1.
These sets of states are also referred as the \emph{almost-sure} winning 
states for the player and an optimal strategy from the 
almost-sure winning states is referred as an almost-sure winning
strategy.
The set $W_1^G(\Phi)$, for B\"uchi or coB\"uchi objectives $\Phi$,
for an MDP $G$ can be computed in $O(n^{\frac{3}{2}})$ time, where 
$n$ is the size of the MDP $G$~\cite{CJH03}.

\medskip\noindent{\bf Attractor of probabilistic states.}
We define a notion of \emph{attractor} of probabilistic states:
given an MDP $G$ and a set $U \subseteq S$ of states,
we denote by $\attr_P(U,G)$ the set of states from where 
the probabilistic player has a strategy (with proper choice
of edges) to force the game to reach $U$.
The set $\attr_P(U,G)$ is inductively defined as follows:
\[
T_0=U; 
\quad 
T_{i+1}=T_i \cup \set{s \in \SR\mid E(s) \cap T_i \neq \emptyset}
\cup \set{s \in S_1 \mid E(s) \subseteq T_i}
\]
and $\attr_P(U,G)=\bigcup_{i \geq 0} T_i$.

We now present a lemma about MDPs with B\"uchi and coB\"uchi objectives
and a property of end components and attractors.
The first two properties of Lemma~\ref{lemm:mdp-prop} follows from 
Lemma~\ref{lemm:end-component1}.
The last property follows from the fact that an end component
is $\trans$-closed (i.e., for an end component $U$, for all
$s \in U \cap \SR$ we have $E(s) \subseteq U$).

\begin{lemma}\label{lemm:mdp-prop}
Let $G$ be an MDP.
Given $B \subseteq S$ and $C \subseteq S$, the following assertions
hold.
\begin{enumerate}
\item For all $U \in \cale(G)$ such that $U \cap B \neq \emptyset$, 
we have $U \subseteq W_1^G(\Buchi{B})$.

\item For all $U \in \cale(G)$ such that $U \subseteq C$, 
we have $U \subseteq W_1^G(\coBuchi{C})$.

\item For all $Y \subseteq S$  and all end components $U \in \cale(G)$,
if $X=\attr_P(Y,G)$, then either (a)~$U \cap Y\neq \emptyset$ or 
(b)~$U \cap X =\emptyset$.
\end{enumerate}
\end{lemma}

\subsection{MDPs with limsup objectives}\label{subsec:limsup}

In this subsection we present polynomial time algorithm for MDPs with
limsup objectives. 
For the sake of simplicity we will consider bipartite MDPs.

\medskip\noindent{\bf Bipartite MDPs.} An MDP $G=((S,E),(S_1,\SR),\trans)$
is \emph{bipartite} if $E \subseteq S_1 \times \SR \cup \SR \times S_1$.
An MDP $G$ can be converted into a bipartite MDP $G'$ by adding dummy
states with an unique successor, and $G'$ is linear in the
size of $G$. 
In sequel without loss of generality we will consider bipartite MDPs.
The key property of bipartite MDPs that will be useful is as follows: 
for a bipartite MDP $G=((S,E),(S_1,\SR),\trans)$, for all $U \in \cale(G)$ 
we have $U \cap S_1 \neq \emptyset$.

\medskip\noindent{\bf Informal description of algorithm.}
We first present an algorithm that takes an MDP $G$ with a 
positive reward function $r:S \to \reals^+$, and computes a
set $S^*$ and a function $f^*:S^* \to \reals^+$.
The output of the algorithm will be useful in reduction of MDPs with
limsup objectives to MDPs with max objectives.
Let the rewards be $v_0 > v_1 >\cdots > v_k$.
The algorithm proceeds in iteration and in iteration $i$ we denote
the MDP as $G_i$ and the state space as $S^i$.
At iteration $i$ the algorithm considers the set $V_i$ of 
reward $v_i$ in the MDP $G_i$, and computes the set  
$U_i=W_1^{G_i}(\Buchi{V_i})$, (i.e., the almost-sure winning
set in the MDP $G_i$ for B\"uchi objective with the 
B\"uchi set $V_i$).
For all $u \in U_i \cap S_i$ we assign $f^*(u)=v_i$ 
and add the set $U_i \cap S_1$ to $S^*$.
Then the set $\attr_P(U_i,G_i)$ is removed from the 
MDP $G_i$ and we proceed to iteration $i+1$.
In $G_i$ all end components that intersect with reward $v_i$ are
contained in $U_i$ (by Lemma~\ref{lemm:mdp-prop} part (1)), 
and all end components in $S^i \setminus U_i$ do not intersect with 
$\attr_P(U_i,G_i)$ (by Lemma~\ref{lemm:mdp-prop} part(3)).
This gives us the following lemma.

\begin{algorithm}[t]
\caption{MDPLimSup}
\label{algorithm:mdp-limsup}
{
\begin{tabbing}
aa \= aa \= aaa \= aaa \= aaa \= aaa \= aaa \= aaa \kill
\\
\> {\bf Input:}  MDP $G=((S,E),(S_1,\SR),\trans)$, a positive reward function $r:S \to \reals^+$. \\
\>   {\bf Output:} $S^* \subseteq S$ and $f^*:S^* \to \reals^+$ \\

\> 1. Let $r(S)=\set{v_0, v_1, \ldots, v_k}$ with $v_0 > v_1 > \cdots > v_k$; \\
\> 2. $G_0:=G$; $S^*=\emptyset$; \\
\> 3. {\bf for} $i:=0$ to $k$ {\bf do } \{ \\ 
\>\> 3.1 $U_i := W_1^{G_i}(\Buchi{r^{-1}(v_i) \cap S^i})$; \\
\>\> 3.2 {\bf for} all $u \in U_i \cap S_1$ \\
\>\>\> $f^*(u) := v_i$; \\
\>\> 3.3 $S^*:=S^* \cup (U_i \cap S_1)$; \\
\>\> 3.4 $B_i:=\attr_P(U_i,G_i)$; \\
\>\> 3.5 $G_{i+1}:=G_i \setminus B_i$, $S^{i+1}:=S^i \setminus B_i$; \\ 
\> \}  \\

\> 4. {\bf return} $S^*$ and $f^*$.
\end{tabbing}
}\end{algorithm}

\begin{lemma}\label{lemm:mdp-limsup1}
Let $G$ be an MDP with a positive reward function $r:S \to \reals^+$.
Let $f^*$ be the output of Algorithm~\ref{algorithm:mdp-limsup}.
For all end components $U\in \cale(G)$ and all states $u\in U \cap S_1$, 
we have $\max(r(U)) \leq f^*(u)$.
\end{lemma}
\begin{proof}
Let $U^*=\bigcup_{i=0}^k U_i$ (as computed in Algorithm~\ref{algorithm:mdp-limsup}).
Then it follows from Lemma~\ref{lemm:mdp-prop} that 
for all $A \in \cale(G)$ we have $A \cap U^* \neq \emptyset$.
Consider $A \in \cale(G)$ and let $v_i=\max(r(A))$.
Suppose for some $j<i$ we have $A \cap U_j \neq \emptyset$.
Then there is a strategy to ensure that $U_j$ is reached with 
probability~1 from all states in $A$ and then play an almost-sure winning 
strategy in $U_j$
to ensure $\Buchi{r^{-1}(v_j) \cap S^j}$.
Then $A \subseteq U_j$. 
Hence for all $u \in A \cap S_1$ we have  $f^*(u)=v_j \geq v_i$.
If for all $j < i$ we have $A \cap U_j =\emptyset$, then 
we show that $A \subseteq U_i$.
The uniform memoryless strategy $\straa_A$ 
(as used in Lemma~\ref{lemm:end-component1})
in $G_i$ is a witness 
to prove that $A \subseteq U_i$.
In this case for all $u \in A \cap S_1$ we have $f^*(u)=v_i
=\max(r(A))$.
The desired result follows.
\qed
\end{proof}

\medskip\noindent{\bf Transformation to MDPs with $\max$ objective.}
Given an MDP $G=((S,E),(S_1,\SR),\trans)$ with a positive 
reward function $r:S\to \reals^+$, and let 
$S^*$ and $f^*$ be the output of Algorithm~\ref{algorithm:mdp-limsup}.
We construct an MDP $\ov{G}=((\ov{S},\ov{E}),(\ov{S}_1,\ovSR),\ov{\trans})$
with a reward function $\ov{r}$ as follows:
\begin{itemize}
\item $\ov{S} =S \cup \wh{S}^*$; i.e., the set of states consists of the 
state space $S$ and a copy $\wh{S}^*$ of $S^*$.
\item $\ov{E}= E \cup \set{(s,\wh{s}) \mid s \in S^*, \wh{s}\in \wh{S}^* \mbox{ where $\wh{s}$ is the copy of $s$}}
	\cup \set{(\wh{s},\wh{s}) \mid \wh{s} \in \wh{S}^*}$; i.e., 
	along with edges $E$, for all states $s$ in $S^*$ there is an 
	edge to its copy $\wh{s}$ in $\wh{S}^*$, and
	all states in $\wh{S}^*$ are absorbing states.

\item $\ov{S}_1=S_1 \cup \wh{S}^*$.

\item $\ov{\trans}=\trans$.

\item $\ov{r}(s)=0$ for all $s \in S$ and $\ov{r}(s)=f^*(s)$ for $\wh{s} \in
\wh{S}^*$, where $\wh{s}$ is the copy of $s$.
\end{itemize}
We refer to the above construction as \emph{limsup} conversion.
The following lemma proves the relationship between the value
function $\va^G(\limsup(r))$ and $\va^{\ov{G}}(\max(\ov{r}))$.

\begin{lemma}
Let $G$ be an MDP with a positive reward function
$r:S\to \reals^+$. Let $\ov{G}$ and $\ov{r}$ be obtained from $G$ and $r$ 
by the limsup conversion.
For all states $s \in S$, we have
\[
\va^G(\limsup(r))(s) =\va^{\ov{G}}(\max(\ov{r}))(s).
\] 
\end{lemma}
\begin{proof}
The result is obtained from the following two case analysis.
\begin{enumerate}

\item Let $\straa$ be a pure memoryless optimal strategy in $G$
for the objective $\limsup(r)$.
Let $\calc=\set{C_1,C_2,\ldots,C_m}$ be the set of closed
connected recurrent sets in the Markov chain obtained from $G$
after fixing the strategy $\straa$.
Note that since we consider bipartite MDPs, for all $1 \leq i \leq m$, 
we have $C_i \cap S_1 \neq \emptyset$.
Let $C=\bigcup_{i=1}^m C_i$.
We define a pure memoryless strategy $\ov{\straa}$ in $\ov{G}$ as follows
\[
\ov{\straa}(s) =
\begin{cases}
\straa(s) & s \in S_1 \setminus C; \\
\wh{s}	& \wh{s} \in \wh{S}^* \mbox{ and } s \in S_1 \cap C.
\end{cases}
\]
By Lemma~\ref{lemm:mdp-limsup1} it follows that the strategy 
$\ov{\straa}$ ensures that for all $C_i \in \calc$ and all 
$s \in C_i$, the maximal reward reached in $\ov{G}$ is at least
$\max(r(C_i))$ with probability~1.
It follows that for all $s \in S$ we have 
\[
\va^G(\limsup(r))(s) \leq\va^{\ov{G}}(\max(\ov{r}))(s).
\]

\item Let $\ov{\straa}$ be a pure memoryless optimal strategy 
for the objective $\max(\ov{r})$ in $\ov{G}$.
We fix a strategy $\straa$ in $G$ as follows:
if at a state $s \in S^*$ the strategy $\ov{\straa}$ chooses 
the edge $(s,\wh{s})$, then in $G$ on reaching $s$, the strategy 
$\straa$ plays an almost-sure winning strategy for the objective
$\Buchi{r^{-1}(f^*(s))}$, otherwise $\straa$ follows $\ov{\straa}$.
It follows that for all $s \in S$ we have 
\[
\va^G(\limsup(r))(s) \geq\va^{\ov{G}}(\max(\ov{r}))(s).
\]
\end{enumerate} 
Thus we have the desired result.
\qed
\end{proof}

\medskip\noindent{\bf Linear programming for the $\max$ objective in $\ov{G}$.}
The following linear program characterizes the value function
$\va^{\ov{G}}(\max(\ov{r}))$.
For all $s \in \ov{S}$ we have a variable $x_s$ and the objective
function is $\min \sum_{s \in \ov{S}} x_s$.
The set of linear constraints are as follows:
\[
\begin{array}{rcll}
x_s &\geq &  0 & \quad \forall s\in \ov{S}; \\
x_s & = & \ov{r}(s) & \quad \forall s \in \wh{S}^*; \\
x_s & \geq & x_t & \quad \forall s \in \ov{S}_1, (s,t) \in \ov{E}; \\
x_s & = & \sum_{t \in \ov{S}} \ov{\trans}(s)(t) \cdot x_t & \quad \forall s\in \ovSR.
\end{array} 
\]
The correctness proof of the above linear program to 
characterize the value function 
$\va^{\ov{G}}(\max(\ov{r}))$ follows by extending the result 
for reachability objectives~\cite{FV97}.
The key property that can be used to prove the correctness of the 
above claim is as follows:
if a pure memoryless optimal strategy is fixed, then from all states in 
$S$, the set $\wh{S}^*$ of absorbing states is reached with probability~1.
The above property can be proved as follows:
since $r$ is a positive reward function, it follows that for all 
$s \in S$ we have $\va^G(\limsup(r))(s)>0$.
Moreover, for all states $s \in S$ we have $\va^{\ov{G}}(\max(\ov{r}))(s)
=\va^G(\limsup(r))(s)>0$.
Observe that for all $s \in S$ we have $\ov{r}(s)=0$.
Hence if we fix a pure memoryless optimal strategy $\straa$ in $\ov{G}$, 
then in the Markov chain $\ov{G}_{\straa}$ there is no closed recurrent set $C$
such that $C \subseteq S$. 
It follows that for all states $s\in S$, in the Markov $\ov{G}_{\straa}$, the
set $\wh{S}^*$ is reached with probability~1. 
Using the above fact and the correctness of linear-programming for 
reachability objectives, the correctness proof of the above linear-program for 
the objective $\max(\ov{r})$ in $\ov{G}$ can be obtained.
This shows that the value function 
$\va^{G}(\limsup(r))$ for MDPs with reward function $r$ can be 
computed in polynomial time. 
This gives us the following result.

\begin{theorem}\label{thrm:limsup}
Given an MDP $G$ with a reward function $r$, the
value function $\va^{G}(\limsup(r))$ can be computed
in polynomial time.
\end{theorem}

\subsection{MDPs with liminf objectives}
In this subsection we present polynomial time algorithms for MDPs with 
liminf objectives, and then present the complexity result for $2\half$-player
games with limsup and liminf objectives.

\medskip\noindent{\bf Informal description of algorithm.}
We first present an algorithm that takes an MDP $G$ with a 
positive reward function $r:S \to \reals^+$, and computes a
set $S_*$ and a function $f_*:S_* \to \reals^+$.
The output of the algorithm will be useful in reduction of MDPs with
liminf objectives to MDPs with max objectives.
Let the rewards be $v_0 > v_1 >\cdots > v_k$.
The algorithm proceeds in iteration and in iteration $i$ we denote
the MDP as $G_i$ and the state space as $S^i$.
At iteration $i$ the algorithm considers the set $V_i$ of 
reward at least $v_i$ in the MDP $G_i$, and computes the set  
$U_i=W_1^{G_i}(\coBuchi{V_i})$, (i.e., the almost-sure winning
set in the MDP $G_i$ for coB\"uchi objective with the 
coB\"uchi set $V_i$).
For all $u \in U_i \cap S_i$ we assign $f_*(u)=v_i$ 
and add the set $U_i \cap S_1$ to $S_*$.
Then the set $\attr_P(U_i,G_i)$ is removed from the 
MDP $G_i$ and we proceed to iteration $i+1$.
In $G_i$ all end components that contain reward at least $v_i$ are
contained in $U_i$ (by Lemma~\ref{lemm:mdp-prop} part (2)), 
and all end components in $S^i \setminus U_i$ do not intersect with 
$\attr_P(U_i,G_i)$ (by Lemma~\ref{lemm:mdp-prop} part(3)).
This gives us the following lemma.

\begin{algorithm}[t]
\caption{MDPLimInf}
\label{algorithm:mdp-liminf}
{
\begin{tabbing}
aa \= aa \= aaa \= aaa \= aaa \= aaa \= aaa \= aaa \kill
\\
\> {\bf Input:}  MDP $G=((S,E),(S_1,\SR),\trans)$, a positive reward function $r:S \to \reals^+$. \\
\>   {\bf Output:} $S_* \subseteq S$ and $f_*:S_* \to \reals^+$. \\

\> 1. Let $r(S)=\set{v_0, v_1, \ldots, v_k}$ with $v_0 > v_1 > \cdots > v_k$; \\
\> 2. $G_0:=G$; $S_*=\emptyset$; \\
\> 3. {\bf for} $i:=0$ to $k$ {\bf do } \{ \\ 
\>\> 3.1 $U_i := W_1^{G_i}(\coBuchi{\bigcup_{j\leq i} r^{-1}(v_j) \cap S^i})$; \\
\>\> 3.2 {\bf for} all $u \in U_i \cap S_1$ \\
\>\>\> $f_*(u) := v_i$; \\
\>\> 3.3 $S_*:=S_* \cup (U_i \cap S_1)$; \\
\>\> 3.4 $B_i:=\attr_P(U_i,G_i)$; \\
\>\> 3.5 $G_{i+1}:=G_i \setminus B_i$, $S^{i+1}:=S^i \setminus B_i$; \\ 
\> \}  \\

\> 4. {\bf return} $S_*$ and $f_*$.
\end{tabbing}
}\end{algorithm}

\begin{lemma}\label{lemm:mdp-liminf1}
Let $G$ be an MDP with a positive reward function $r:S \to \reals^+$.
Let $f_*$ be the output of Algorithm~\ref{algorithm:mdp-liminf}.
For all end components $U\in \cale(G)$ and all states $u\in U \cap S_1$, 
we have $\min(r(U)) \leq f^*(u)$.
\end{lemma}
\begin{proof}
Let $U^*=\bigcup_{i=0}^k U_i$ (as computed in Algorithm~\ref{algorithm:mdp-liminf}).
Then it follows from Lemma~\ref{lemm:mdp-prop} that 
for all $A \in \cale(G)$ we have $A \cap U^* \neq \emptyset$.
Consider $A \in \cale(G)$ and let $v_i=\min(r(A))$.
Suppose for some $j<i$ we have $A \cap U_j \neq \emptyset$.
Then there is a strategy to ensure that $U_j$ is reached with 
probability~1 from all states in $A$ and then play an almost-sure winning 
strategy in $U_j$
to ensure $\coBuchi{\bigcup_{l \leq j} r^{-1}(v_l) \cap S^j}$.
Then $A \subseteq U_j$. 
Hence for all $u \in A \cap S_1$ we have  $f_*(u)=v_j \geq v_i$.
If for all $j < i$ we have $A \cap U_j =\emptyset$, then 
we show that $A \subseteq U_i$.
The uniform memoryless strategy $\straa_A$ 
(as used in Lemma~\ref{lemm:end-component1}) 
in $G_i$ is a witness 
to prove that $A \subseteq U_i$.
In this case for all $u \in A \cap S_1$ we have $f_*(u)=v_i
=\min(r(A))$.
The desired result follows.
\qed
\end{proof}

\medskip\noindent{\bf Transformation to MDPs with $\max$ objective.}
Given an MDP $G=((S,E),(S_1,\SR),\trans)$ with a positive 
reward function $r:S\to \reals^+$, and let 
$S_*$ and $f_*$ be the output of Algorithm~\ref{algorithm:mdp-liminf}.
We construct an MDP $\ov{G}=((\ov{S},\ov{E}),(\ov{S}_1,\ovSR),\ov{\trans})$
with a reward function $\ov{r}$ as follows:
\begin{itemize}
\item $\ov{S} =S \cup \wh{S}_*$; i.e., the set of states consists of the 
state space $S$ and a copy $\wh{S}_*$ of $S_*$.
\item $\ov{E}= E \cup \set{(s,\wh{s}) \mid s \in S_*, \wh{s}\in \wh{S}_* \mbox{ where $\wh{s}$ is the copy of $s$}}
	\cup \set{(\wh{s},\wh{s}) \mid \wh{s} \in \wh{S}_*}$;
	along with edges $E$, for all states $s$ in $S_*$ there is an 
	edge to its copy $\wh{s}$ in $\wh{S}_*$, and
	all states in $\wh{S}_*$ are absorbing states.

\item $\ov{S}_1=S_1 \cup \wh{S}_*$.

\item $\ov{\trans}=\trans$.

\item $\ov{r}(s)=0$ for all $s \in S$ and $\ov{r}(s)=f_*(s)$ for $\wh{s} \in
\wh{S}_*$, where $\wh{s}$ is the copy of $s$.
\end{itemize}
We refer to the above construction as \emph{liminf} conversion.
The following lemma proves the relationship between the value
function $\va^G(\liminf(r))$ and $\va^{\ov{G}}(\max(\ov{r}))$.

\begin{lemma}
Let $G$ be an MDP with a positive reward function
$r:S\to \reals^+$. Let $\ov{G}$ and $\ov{r}$ be obtained from $G$ and $r$ 
by the liminf conversion.
For all states $s \in S$, we have
\[
\va^G(\liminf(r))(s) =\va^{\ov{G}}(\max(\ov{r}))(s).
\] 
\end{lemma}
\begin{proof}
The result is obtained from the following two case analysis.
\begin{enumerate}

\item Let $\straa$ be a pure memoryless optimal strategy in $G$
for the objective $\liminf(r)$.
Let $\calc=\set{C_1,C_2,\ldots,C_m}$ be the set of closed
connected recurrent sets in the Markov chain obtained from $G$
after fixing the strategy $\straa$.
Since $G$ is an bipartite MDP, it follows that for all $1 \leq i \leq m$,
we have $C_i \cap S_1 \neq \emptyset$.
Let $C=\bigcup_{i=1}^m C_i$.
We define a pure memoryless strategy $\ov{\straa}$ in $\ov{G}$ as follows
\[
\ov{\straa}(s) =
\begin{cases}
\straa(s) & s \in S_1 \setminus C; \\
\wh{s}	& \wh{s} \in \wh{S}_* \mbox{ and } s \in S_1 \cap C.
\end{cases}
\]
By Lemma~\ref{lemm:mdp-liminf1} it follows that the strategy 
$\ov{\straa}$ ensures that for all $C_i \in \calc$ and all 
$s \in C_i$, the maximal reward reached in $\ov{G}$ is at least
$\min(r(C_i))$ with probability~1.
It follows that for all $s \in S$ we have 
\[
\va^G(\limsup(r))(s) \leq\va^{\ov{G}}(\max(\ov{r}))(s).
\]

\item Let $\ov{\straa}$ be a pure memoryless optimal strategy 
for the objective $\max(\ov{r})$ in $\ov{G}$.
We fix a strategy $\straa$ in $G$ as follows:
if at a state $s \in S_*$ the strategy $\ov{\straa}$ chooses 
the edge $(s,\wh{s})$, then in $G$ on reaching $s$, the strategy 
$\straa$ plays an almost-sure winning strategy for the objective
$\coBuchi{\bigcup_{v_j \geq f_*(s)} r^{-1}(v_j)}$, otherwise $\straa$ follows $\ov{\straa}$.
It follows that for all $s \in S$ we have
\[
\va^G(\liminf(r))(s) \geq\va^{\ov{G}}(\max(\ov{r}))(s).
\]
\end{enumerate} 
Thus we have the desired result.
\qed
\end{proof}

\medskip\noindent{\bf Linear programming for the $\max$ objective in $\ov{G}$.}
The linear program of subsection~\ref{subsec:limsup} characterizes the value 
function $\va^{\ov{G}}(\max(\ov{r}))$.
This shows that the value function 
$\va^{G}(\liminf(r))$ for MDPs with reward function $r$ can be 
computed in polynomial time. 
This gives us the following result.

\begin{theorem}\label{thrm:liminf}
Given an MDP $G$ with a reward function $r$, the
value function $\va^{G}(\liminf(r))$ can be computed
in polynomial time.
\end{theorem}

\subsection{$2\half$-player games with limsup and liminf objectives}
We now show that $2\half$-player games with limsup and liminf 
objectives can be decided in NP $\cap$ coNP.
The pure memoryless optimal strategies (existence 
follows from Theorem~\ref{thrm:quan-det})
provide the polynomial witnesses and to obtain the 
desired result we need to present a polynomial time
verification procedure.
In other words, we need to present polynomial time algorithms for 
MDPs with limsup and liminf objectives. 
Since the value functions in MDPs with limsup and liminf objectives
can be computed in polynomial time 
(Theorem~\ref{thrm:limsup} and Theorem~\ref{thrm:liminf}),
we obtain the following result about the complexity 
$2\half$-player games with limsup and liminf objectives.

\begin{theorem}
Given a $2\half$-player game graph $G$ with a reward 
function $r$, a state $s$ and a rational value $q$,
the following assertions hold:
(a) whether $\va^G(\limsup(r))(s) \geq q$ can be decided in
NP $\cap$ coNP; and
(b) whether $\va^G(\liminf(r))(s) \geq q$ can be decided in
NP $\cap$ coNP.
\end{theorem}

\medskip\noindent{\bf Acknowledgments.}
{\small We thank Hugo Gimbert for explaining his results and pointing out
relevant literature on games with limsup and liminf objectives.
This research was supported in part by the NSF grants
CCR-0132780, CNS-0720884, and CCR-0225610, by the Swiss
National Science Foundation, and by the COMBEST project of the European Union.
}

\bibliographystyle{plain}
\bibliography{main}

\begin{thebibliography}{10}

\bibitem{ChaHen07b}
K.~Chatterjee and T.A. Henzinger.
\newblock Value iteration.
\newblock In {\em 25 Years of Model Checking}, LNCS. Springer, 2007.

\bibitem{CJH03}
K.~Chatterjee, M.~Jurdzi{\'n}ski, and T.A. Henzinger.
\newblock Simple stochastic parity games.
\newblock In {\em CSL'03}, volume 2803 of {\em LNCS}, pages 100--113. Springer,
  2003.

\bibitem{CJH04}
K.~Chatterjee, M.~Jurdzi{\'{n}}ski, and T.A. Henzinger.
\newblock Quantitative stochastic parity games.
\newblock In {\em SODA'04}, pages 121--130. SIAM, 2004.

\bibitem{CY90}
C.~Courcoubetis and M.~Yannakakis.
\newblock {Markov} decision processes and regular events.
\newblock In {\em ICALP 90: Automata, Languages, and Programming}, volume 443
  of {\em Lecture Notes in Computer Science}, pages 336--349. Springer-Verlag,
  1990.

\bibitem{luca-Thesis}
L.~de~Alfaro.
\newblock {\em Formal Verification of Probabilistic Systems}.
\newblock PhD thesis, Stanford University, 1997.

\bibitem{FV97}
J.~Filar and K.~Vrieze.
\newblock {\em Competitive {Markov} Decision Processes}.
\newblock Springer-Verlag, 1997.

\bibitem{Gimbert06}
H.~Gimbert.
\newblock {\em Jeux positionnels}.
\newblock PhD thesis, Universit\'e Paris~7, 2006.

\bibitem{GZ05}
H.~Gimbert and W.~Zielonka.
\newblock Games where you can play optimally without any memory.
\newblock In {\em CONCUR'05}, pages 428--442. Springer, 2005.

\bibitem{GimZie}
H.~Gimbert and W.~Zielonka.
\newblock Perfect information stochastic priority games.
\newblock In {\em ICALP'07}, pages 850--861. Springer, 2007.

\bibitem{LigLip69}
T.~A. Liggett and S.~A. Lippman.
\newblock Stochastic games with perfect information and time average payoff.
\newblock {\em Siam Review}, 11:604--607, 1969.

\bibitem{MaitraSudderth}
A.~Maitra and W.~Sudderth, editors.
\newblock {\em Discrete Gambling and Stochastic Games}.
\newblock Springer, 1996.

\bibitem{Mar98}
D.A. Martin.
\newblock The determinacy of {Blackwell} games.
\newblock {\em The Journal of Symbolic Logic}, 63(4):1565--1581, 1998.

\end{thebibliography}

\end{document}